\documentclass[twocolumns]{aa}

\usepackage{lineno,hyperref}

\usepackage{natbib}
\usepackage{amssymb}
\usepackage{subfig}
\usepackage{graphicx}

\usepackage{amsmath}	
\usepackage[dvipsnames]{xcolor} 
\usepackage[T1]{fontenc}
\usepackage{ae,aecompl}
\usepackage[english]{babel}

\usepackage{hyperref}
\usepackage{url}
\usepackage{soul}

\usepackage{graphicx}	
\usepackage{amsmath}	
\usepackage{amssymb}	
\usepackage{adjustbox}
\usepackage{threeparttable} 
\usepackage[normalem]{ulem}

\newcommand{\be}{\begin{equation}}
\newcommand{\ee}{\end{equation}}
\newcommand{\bea}{\begin{eqnarray}}

\newcommand{\eea}{\end{eqnarray}}
\DeclareMathAlphabet\mathbfcal{OMS}{cmsy}{b}{n}

\hypersetup{colorlinks=true, citecolor=blue, linkcolor=blue}

\modulolinenumbers[5]

\begin{document}

\title{Semi-analytical model for planetary resonances:}
\subtitle{application to planets around single and binary stars}

   \author{Tabar\'e Gallardo
          \inst{1}\fnmsep\thanks{E-mail: gallardo@fisica.edu.uy},
          Cristián Beaugé\inst{2}, Cristian A. Giuppone\inst{2}
          }

   \institute{Instituto de F\'{i}sica, Facultad de Ciencias, Udelar, Igu\'{a} 4225, 11400 Montevideo, Uruguay
        \and
        Universidad Nacional de C\'ordoba, Observatorio Astron\'omico - IATE. Laprida 854, 5000 C\'ordoba, Argentina
         }

\date{}

\abstract
{In spite of planetary resonances being a common dynamical mechanism acting on planetary systems, no general model exists for describing their properties, particularly for commensurabilities of any order and arbitrary values of the eccentricities and inclinations.}
{The present work presents a semi-analytical model that describes the resonance strength, width, location and stability of fixed points, as well as periods of small-amplitude librations. The model is valid to any two gravitationally interacting massive bodies, and thus applicable to planets around single or binary stars.}
 {Using a theoretical framework in Poincaré and Jacobi reference system we develop a semi-analytical method that employs a numerical evaluation of the averaged resonant disturbing function. Validations of the model are presented comparing its predictions with dynamical maps for real and fictitious systems.} 
{The model is shown to describe very well many dynamical features of planetary resonances. Notwithstanding the good agreements found in all cases, a small deviation is noted in the location of the resonance centers for circumbinary systems. As a consequence of its application to the HD31527 system we have found that the updated best-fit solution leads to a high-eccentricity stable libration between the middle and outer planets inside the 16/3 mean-motion resonance. This is the first planetary system whose long-term dynamics appears dominated by such a high-order commensurability. In the case of circumbinary planets, the overlap of N/1 mean-motion resonances coincides very well with the size of the global chaotic region close to the binary, as well as its dependence with the mutual inclination.}
{}


\titlerunning{Semi-analytical model for planetary resonances}
\authorrunning{Gallardo et al.}
\maketitle


\section{Introduction}
\label{intro}

There is a fundamental difference between asteroidal and planetary resonances: in the first case
all the dynamics is restricted to the particle while in the second case both massive bodies
participate and experience the dynamical effects of the resonance. 
In the asteroidal (or restricted) case  the planet does not perceive the existence of the resonance. The asteroid or particle can evolve in a resonance interior to the planet orbit or exterior to it or could be coorbital with the planet without perturbing the planet.
In planetary resonances we do not distinguish between interior or exterior resonances because both massive bodies define the dynamics. 
Both massive bodies librate with the same period but 
it is intuitive that the dynamical effects of the resonance on each planet will be inversely proportional to their masses, and when one of them has a negligible mass the restricted asteroidal case is reproduced.

As first order resonances play a fundamental role in the structure of planetary systems most attention has been deserved to this type of resonances.
There are analytical models that describe very well the resonances between two planets for coplanar orbits \citep{Beauge2003,Batygin2013b,2013ApJ...774..129D} and there are several studies regarding the more complex problem of three body planetary resonances \citep[e.g.][]{Quillen2011,Gallardo2016a,2017A&A...605A..96D}
but no model exists capable of handling arbitrary two-body resonances between massive bodies with arbitrary orbits.

To attack the problem at least three points must be defined: the system of variables, the dimensions of the problem (planar or spatial) and the way the resonant disturbing function is computed. The astrocentric system can be adopted but we cannot obtain a set of canonical variables from it, so Poincaré or Jacobi systems should be adopted if we look for a canonical framework. The planar case of course is simpler and is the rule for analytical treatments. The spatial case adds complexity and it was only studied by numerical methods \citep{2013CeMDA.115..161A}. Regarding the resonant disturbing function, it is obtained analytically from general classic Laplacian developments \citep{Batygin2013b} or from developments for the planar case with improved convergence domain \citep{Beauge2003}.

The most important difference of our model with respect to other ones is that 
we calculate numerically the resonant disturbing function, no expansions are involved. In section \ref{Poincaré} we explain the details of this calculation. In order to simplify the system of equations we also assume that during the resonant motion 
 both orbits have fixed longitudes of the nodes and periastron. 
 This is well justified when the time scale of the resonant motion is much smaller than the time scale of secular evolution. 
 This approach reduce the involved angular variables to the mean longitudes and the resulting critical angle. With this hypothesis we can numerically calculate the resonant disturbing function and to obtain the properties of the resonance: strength, width, equilibrium points and period of the small amplitude oscillations. 
This approach is a generalization of the model given by \cite{2020CeMDA.132....9G} for the restricted case, it reproduces very well
the resonant mode but no information can be obtained related to the long term behaviour of the system.
Closely related to this problem is the study of the dynamics of a planet in resonance with a star in a binary system \citep{2012MNRAS.424...52M}. This case is also contemplated by our method.

Following, in section \ref{Poincaré} we develop the semi-analytical model in Poincaré elements, which is suitable for a system with a star and two planets. Then, we move to the section \ref{jacobi}, where we present the semi-analytical model in Jacobi elements, which is more general and in particular suitable for a circumbinary planet. In section \ref{sec.planets} we apply the model to describe the resonance between pair of planets in HD31527 and HD74156, while in section \ref{sec.cbp} we apply the model in Jacobi coordinates to explain chaotic regions for planets around a binary. Conclusions close the paper in section \ref{sec.conclusions}.

\section{Resonance model in Poincaré coordinates}
\label{Poincaré}
Astrocentric or baricentric coordinates are not canonical when applied to a planetary system. In this case the Poincaré coordinates are a valid alternative. 
According to  \cite{Laskar.1991, Laskar+Robutel.1995, Ferraz-Mello.2006}
the Hamiltonian for the system of a star with mass $m_0$ and two planets $m_1$ and $m_2$ in Poincaré coordinates is given by
\begin{equation}\label{h}
H= -\frac{\mu_1^2 \beta_1^3}{2L_1^2} - \frac{\mu_2^2 \beta_2^3}{2L_2^2} - R
\end{equation}
where
\begin{equation}\label{r}
R= k^2\frac{m_1m_2}{\Delta} - \frac{m_1m_2}{m_0}\vec{v_1}\cdot\vec{v_2}
\end{equation}
with $k$ being the Gaussian constant, $\vec{v_1}$ and $\vec{v_2}$ the barycentric velocity vectors, $\mu_i=k^2(m_0+m_i)$,
$\beta_i=m_0m_i/(m_0+m_i)$, $L_i=\beta_i\sqrt{\mu_i a_i}$, $\Delta$ is the mutual distance of the planets and being $a_i$ the semimajor axes defined in the Poincaré system, that means astrocentric positions and barycentric velocities.
The canonical variables are

\begin{equation}\label{canvar}
\begin{array}{ll}
\lambda_i, & L_i = \beta_i\sqrt{\mu_i a_i} \\
\varpi_i, & G_i-L_i = L_i (\sqrt{1-e_i^2}-1)\\
\Omega_i, & H_i-G_i = G_i(\cos i_i -1).\\
\end{array}
\end{equation}

Now consider the planets are close to the resonance $k_2/k_1$ so that $k_1 n_1 \simeq k_2 n_2$ (being $n_i=\sqrt{\mu_i/a^3_i}$) and the combination $k_1 \lambda_1 - k_2 \lambda_2$ is a slow varying variable. 
Note that $k_2\geq k_1$.
We will assume that during a libration period the variables $\varpi_i,\Omega_i$ are constants, then  $(G_i-L_i)$ and $(H_i-G_i)$ become constant parameters, not variables,
so we reduce to two degrees of freedom. 
This is a strong simplification that greatly reduces the complexity of the problem. 
However, it is an acceptable approximation since, in general, the time scale in which the resonant movement occurs is much shorter than the periods of circulation of nodes and pericenters.
In order to isolate the slow varying angle we perform the canonical transformation

\begin{equation}\label{tc}
\begin{array}{ll}
\theta=k_1\lambda_1-k_2\lambda_2, & I_1=L_1/k_1 \\
\lambda_2, & I_2=L_2+k_2L_1/k_1 \
\end{array}
\end{equation}
then the new Hamiltonian is $H(\theta,\lambda_2,I_1,I_2)=H_0(I_1,I_2)-R(\theta,\lambda_2,I_1,I_2)$
with
\begin{equation}\label{hcero}
H_0(I_1,I_2)=-\frac{\mu_1^2 \beta_1^3}{2(k_1I_1)^2} - \frac{\mu_2^2 \beta_2^3}{2(I_2-k_2I_1)^2} 
.
\end{equation}

Now we perform a numerical averaging eliminating $\lambda_2$, obtaining a new Hamiltonian $H(\theta,I_1;I_2)$ with $I_2=$ constant. In order to perform this averaging we assume the simplification of \cite{2020CeMDA.132....9G} where the mean of $R(\theta,\lambda_2,I_1,I_2)$ with respect to $\lambda_2$ is calculated numerically assuming fixed values of $(\theta,I_1,I_2)$. This assumption implies that the orbital elements are fixed during the computation of the numerical averaging that spans over $k_1$ orbital periods of the exterior planet or equivalently
 $k_2$ orbital periods of the interior planet.

The numerical calculation is done as follows. We fix $\theta=0\degr$ and varying $\lambda_2$ from $0$ to $k_1 2\pi$ we calculate the corresponding $\lambda_1=(\theta+k_2\lambda_2)/k_1$. Given both mean longitudes we calculate the rectangular coordinates and velocities for both bodies  and using (\ref{r}) we calculate $R$. We repeat the calculation of $R$ for $1000 k_2$ equispaced values of $\lambda_2$ between in the interval $(0,k_1 2\pi)$. Computing the mean of the calculated $R$ we obtain $R(0\degr)$ which is equivalent to a numerical integration using the rectangle rule.    Now we repeat the procedure for $\theta=1\degr$ and so on with a series of values $\theta_i$ up to obtain a numerical representation $R(\theta)$ with $0\degr\leq \theta \leq 360\degr$.  The last simplification that we adopt is that $R(\theta,I_{1})$ is evaluated at the exact resonance given by $I_{1}=I_0$ then	

\begin{equation}\label{hfin}
H(\theta,I_1;I_2)\simeq H_0(I_1;I_2)-R(\theta,I_{1}=I_0; I_2) .
\end{equation}
Taking into account that $I_1$ in $R$ is evaluated at the exact resonance  $(I_1=I_0)$ and remembering that $I_2$ is constant, then we have $H_0$ depending only on the variable $I_1$ and $R$ depending only on the variable $\theta$, remaining $I_2$ as a parameter. It is possible to calculate the level curves of this Hamiltonian in the space $(\theta,I_1)$ where the separatrix and equilibrium points can be visualized. 
The equilibrium points are defined by $dI_1/dt=d\theta/dt=0$ which implies
\begin{equation}\label{equil}
\frac{dH_0}{dI_1}=0, \frac{dR}{d\theta}=0 .
\end{equation}
The first equation conducts to the trivial result $n_1k_1=n_2k_2$ 
(exact resonance),
and the second one can be computed numerically giving the values of the equilibrium points $\theta_0$. It is standard to refer the equilibrium points in terms of  the critical angle 
\begin{equation}\label{sigma}
\sigma= \theta + (k_2-k_1)\varpi_i
\end{equation}
which has a meaning for near coplanar orbits, but in the spatial case several combinations of $\Omega_i$ with $\varpi_i$ can be present in $\sigma$. Moreover, at high eccentricity or inclination regime several critical angles could be relevant for a specific resonance. Our model impose fixed $\Omega_i$ and $\varpi_i$, then our results describing the dynamics of the resonance are independent of these angles.

We drop here the subindex of $I$ for simplicity but remembering that we are referring to the variable  $I_1$.
Being 
$(\theta_0,I_0)$ 
an equilibrium point in canonical variables, if we consider a some small displacement $(\Delta \theta =s, \Delta I =S)$,  a first-order expansion of the Hamiltonian around the equilibrium point can be written as
\begin{equation}
	H(\theta,I) = H(\theta_0,I_0)  + H_{\theta}s + H_{I}S
\end{equation}
where subscripts in  $H$ mean partial derivatives evaluated at the equilibrium point.
Using the canonical equations we can obtain 
\begin{equation}
\frac{dS}{dt} = -H_{\theta}  = - H_{\theta\theta}s - H_{\theta I}S
\end{equation}
\begin{equation}
\frac{ds}{dt} = H_{I}  = H_{I\theta}s + H_{II}S
\end{equation}
with  $H_{I\theta}=H_{\theta I}=0$ because of the separation of variables given in the approximate Hamiltonian (\ref{hfin}).  Looking for solutions of the type $S=A\exp(2\pi t/T)$ and $s=B\exp(2\pi t/T)$
it is straightforward to prove that oscillations only occur with a
libration period, $T$, in years given by
\begin{equation}\label{libper}
T=\frac{2\pi}{\sqrt{H_{II}H_{\theta\theta}}}
\end{equation}
with
\begin{equation}\label{hii}
H_{II}=-\frac{3k_1^2}{\beta_1a_1^2}-\frac{3k_2^2}{\beta_2a_2^2}
\end{equation}
and $H_{\theta \theta}=-R_{\theta \theta} $ which can be calculated numerically
and both  $H_{II},H_{\theta \theta}$ are evaluated at the stable equilibrium point.

\begin{figure}[ht!]
	\centering
	\includegraphics[width=1.\columnwidth,clip]{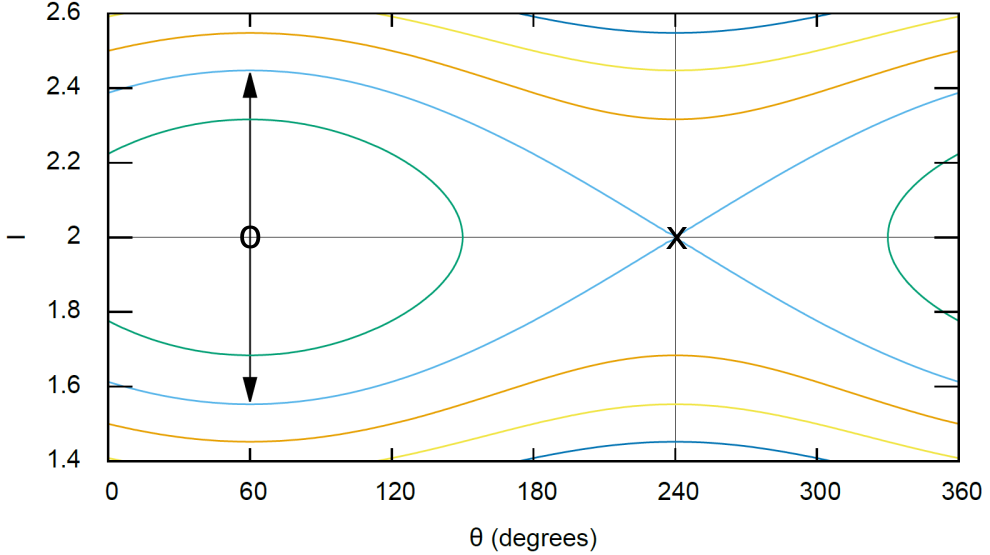}
	\caption{Level curves of a  fictitious Hamiltonian showing with an "O" a stable equilibrium point at $(\theta_s=60\degr, I_0=2)$ 
	and with an "X" an unstable equilibrium point at 
	$(\theta_u=240\degr, I_0=2)$. The arrows show the total resonance's width which is $2 (I_{sep}-I_0)$ being $I_{sep}=2.44$.}
	\label{hamficto}
\end{figure}

To obtain the width of the resonance in terms of $\Delta I$ we follow the same reasoning than in \cite{2020CeMDA.132....9G}. Looking at Fig. \ref{hamficto},
the resonance's half width
is equal to the difference between $I_0$ and $I_{sep}$ where $I_{sep}$ is defined by the separatrix such that
\begin{equation}\label{hh}
H(\theta_s,I_{sep})=H(\theta_u,I_{0})
\end{equation}
being $\theta_s$ and $\theta_u$ the stable and unstable equilibrium points.
Assuming the domain of the resonance is symmetric with respect to the nominal position, $I_0$, 
the total width is twice  $\Delta I$. 
In some cases asymmetries with respect to $I=I_0$ are present and a precise examination of the Hamiltonian level curves is required to obtain the exact width. But in general, being
$\Delta H=H(\theta_s,I_{sep})-H(\theta_s,I_{0})$,  we can approximate
\begin{equation}
\Delta H=\frac{\partial H}{\partial I} \Delta I + 
\frac{\partial^2 H}{\partial I^2} \frac{(\Delta I)^2}{2} + \dots
\end{equation}
Evaluating the derivatives at the stable equilibrium point $(\theta_s,I_0)$ and using Eq. (\ref{hh})
we have
\begin{equation}
\Delta H=H(\theta_u,I_{0})-H(\theta_s,I_{0})\simeq
\frac{\partial^2 H}{\partial I^2} \frac{(\Delta I)^2}{2} .
\end{equation}
Using (\ref{hfin}), the left hand is 
\begin{equation}
H(I_{0})-R(\theta_u)-(H(I_{0})-R(\theta_s))=-(R(\theta_u)-R(\theta_s))=-\Delta R
\end{equation}
while
$\frac{\partial^2 H}{\partial I^2}=H_{II} $
given by  Eq. (\ref{hii}), then

\begin{equation}\label{key}
-\Delta R = {H}_{II} \frac{(\Delta I)^2}{2} .
\end{equation}
Remember that $\Delta I$ is in fact  $\Delta I_1$. 
It is usual to define strength as $SR=<R>-R_{min}$ \citep{2006Icar..184...29G}  which can be  associated with  the coefficient multiplying $\cos(\sigma)$ 
when $R$ is a well behaved cosine-like function, but in general $R$ could be very different from a simple cosine, so in these situations $SR$ gives a reliable estimation of the sensitivity of $R$ with $\sigma$.
The strength is a global property of the resonance, as the libration center and period. It is an indicator of the dynamical relevance of the resonance and it generates different $\Delta a$   on each planet.

The width in terms of $\Delta a_1$ is obtained from the definition of $I_1$ in   Eq.
(\ref{tc})
and $\Delta a_2$ is obtained from $I_2=$ constant and evaluated at the center of the resonance. We obtain the half widths in au:

\begin{equation}\label{dela1}
\Delta a_1=\sqrt{a_1}\frac{\sqrt{m_0+m_1}}{m_0m_1}k_1 2\sqrt{2}\sqrt{\frac{-\Delta R}{k^2 H_{II}}}
\end{equation}
\begin{equation}\label{dela2}
\Delta a_2=\sqrt{a_2}\frac{\sqrt{m_0+m_2}}{m_0m_2}k_2 2\sqrt{2}\sqrt{\frac{-\Delta R}{k^2 H_{II}}}
.
\end{equation}

Note that the larger the planetary mass $m_i$ the narrower the relative width, $\Delta a_i / a_i$. The resonance is the same dynamical mechanism that affect both planets but larger effects are present in the planet with lower mass.

It is important to stress that $\Delta a_i$ are given in terms of Poincaré elements, but for planetary systems the computed widths $\Delta a_i$ in Poincaré elements are very similar to the widths in astrocentric elements. Also, the widths obtained by the formulae above are correct as long as $\Delta a << a$. If large widths are obtained using these formulae it is convenient to recalculate the resonance limits $(I_{max}, I_{min})$ by a numerical procedure looking at the separatrix obtained from the Hamiltonian given by Eq. (\ref{hfin}). In this circumstance it is 
possible to observe asymmetries between 
both limits of the resonance around the exact (nominal) position of the resonance.

\section{Resonance model in Jacobi coordinates}
\label{jacobi}
It is also possible to formulate the method in Jacobi coordinates which are more general than Poincaré coordinates and specially  suited for the study of planets evolving in binary stellar systems.
Following \cite{Lee2003} the Hamiltonian in Jacobi coordinates is

\begin{equation}\label{hjac}
H= -k^4 \frac{m_0^3m_1^3}{2(m_0+m_1)L_1^2} 
-k^4 \frac{(m_0+m_1)^3m_2^3}{2(m_0+m_1+m_2)L_2^2}
- R
\end{equation}
where
\begin{equation}\label{rjaco}
R= k^2m_0m_2(\frac{1}{\Delta_{02}} -\frac{1}{r_{2}} ) +
 k^2m_1m_2(\frac{1}{\Delta_{12}} -\frac{1}{r_{2}} ) 
\end{equation}
where $a_i$ are the osculating semimajor axes in Jacobi system, $\Delta_{i2}$ are the mutual distances between bodies $2$ and $i$ and $r_2$ is the distance of the mass $m_2$ to the barycenter of the subsystem $m_0$ and $m_1$.
The canonical variables are defined in analogy to the Poincaré system but now we have

\begin{equation}\label{l1jaco}
L_1=\frac{m_0m_1}{m_0+m_1}\sqrt{k^2(m_0+m_1)a_1}
\end{equation}
\begin{equation}\label{l2jaco}
L_2=\frac{(m_0+m_1)m_2}{m_0+m_1+m_2}\sqrt{k^2(m_0+m_1+m_2)a_2} .
\end{equation}
We perform the same canonical transformation given by Eq. (\ref{tc}) and
then the new Hamiltonian is $H(\theta,\lambda_2,I_1,I_2)=H_0(I_1,I_2)-R(\theta,\lambda_2,I_1,I_2)$
where
\bea\label{hcerojac}
H_0(I_1,I_2) &=& -k^4 \frac{m_0^3m_1^3}{2(m_0+m_1)k_1^2I_1^2} \nonumber \\ 
             && -k^4 \frac{(m_0+m_1)^3m_2^3}{2(m_0+m_1+m_2)(I_2-k_2I_1)^2} .
\eea
We perform a mean on $\lambda_2$ 
obtaining a Hamiltonian $H(\theta,I_1,I_2)$ analogue to  Eq. (\ref{hfin}). We can calculate level curves for this Hamiltonian and the equilibrium points will show up as well as the separatrix.

\begin{figure}[ht!]
	\centering
	\includegraphics[width=0.99\columnwidth,clip]{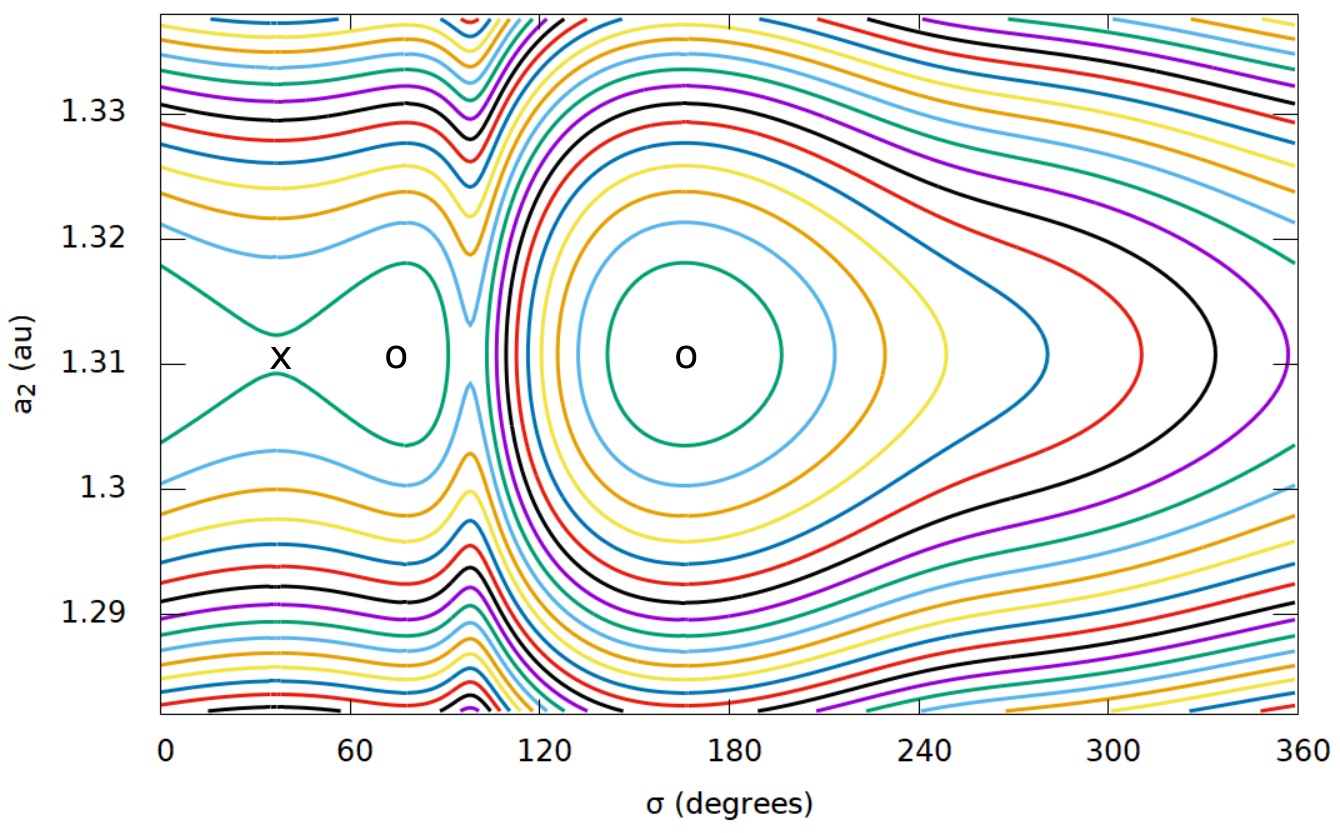}
	\caption{Hamiltonian level curves for a fictitious planetary system composed by a star with  $m_0= 1 M_{\odot}$ and two planets both with mass $0.001 M_{\odot}$, mutual inclination of $20\degr$, $e_1=0.5$ and $e_2=0.9$ located in the resonance 3/2 with $a_1=1$ au.
	 The stable equilibrium points are showed with an "O" and the unstable one is showed with an "X". Using Eq. (\ref{dela2ja}) our model predicts a total width of $0.048$ au in $a_2$.}
	\label{hamtwojup}
\end{figure}

The equilibrium points will verify the same condition given by  Eq. (\ref{equil}).
The first one gives the condition for the exact resonance $k_1n_1=k_2n_2$ where the mean motions are now $n_1=\sqrt{k^2(m_0+m_1)/a_1^3}$ and 
$n_2=\sqrt{k^2(m_0+m_1+m_2)/a_2^3}$.
The second one gives the values of the libration centers $\theta_0$ and the stable libration periods are given by Eq. (\ref{libper}) where in this case
\begin{equation}\label{hiijaco}
H_{II}=-\frac{3k_1^2(m_0+m_1)}{m_0m_1a_1^2}-
\frac{3k_2^2(m_0+m_1+m_2)}{(m_0+m_1)m_2a_2^2}
\end{equation}
and $H_{\theta \theta}=-R_{\theta \theta} $ which is calculated numerically
and both  $H_{II},H_{\theta \theta}$ are evaluated at the stable equilibrium point.
Following an analogue reasoning of the previous section we obtain
\begin{equation}\label{dela1ja}
\Delta a_1 = \sqrt{a_1}\frac{\sqrt{(m_0+m_1)}}{m_0m_1}k_1 2\sqrt{2}\sqrt{\frac{-\Delta R}{k^2H_{II}}}
\end{equation}
\begin{equation}\label{dela2ja}
\Delta a_2 = \sqrt{a_2}\frac{\sqrt{(m_0+m_1+m_2)}}{(m_0+m_1)m_2}k_2 2\sqrt{2}\sqrt{\frac{-\Delta R}{k^2H_{II}}}
.
\end{equation}
We show an example at Fig. \ref{hamtwojup}.
The expressions above give us the half widths of the resonances for the general case. 
For the particular case 
  $m_2/m_1 \sim 0$ which corresponds to  a planet around a binary system,  we obtain $H_{II}\sim -\frac{3k_2^2}{m_2a_2^2}$,
$\Delta a_1 \sim 0$ and  
\begin{equation}
	\Delta a_2 \simeq \frac{\sqrt{8/3}}{n_2}\sqrt{\Delta R /m_2}.
\end{equation}
recovering the resonance width for the restricted case \citep{2020CeMDA.132....9G} in Jacobi coordinates since in the restricted case $m_2$ is not present in $R$.
 In the case of a planet orbiting close to a stellar binary system it is possible that the widths estimated by this method become large in comparison with the planet semimajor axis. In this case a better result can be obtained calculating numerically the borders of the resonance identifying the separatrix using the expression of the Hamiltonian  given by (\ref{hfin}). But no method can compute the resonance widths for those cases where the planet is located so close to the binary so that close encounters occur turning its dynamic chaotic.
 
 To avoid erroneous results generated by non realistic $\Delta R$ due to close encounters, a good practice is to monitoring the mutual distance in the computation of the integral $R(\theta)$ and discard the calculations obtained when close encounters occur. 
 A reasonable tolerance criterion could be around 3 mutual Hill radius, see for example \cite{2020CeMDA.132....9G} and the discussion in the next Section.

Remember that our model assumes fixed pericenters and nodes. 
As they evolve over time, the resonance widths will change over time. For a given set of pericenters and nodes the widths are maximum and for others they are minimum and this leads to the idea of resonance fragility \citep{2020CeMDA.132....9G}, a parameter that measures the variability of the widths.

Results obtained by this model are quite good in comparison with the numerical integration of the full equations of motion and improve the results obtained by analytical methods based in truncated expansions. Looking at the system
HD45364 studied by \cite{Batygin2013b} for example, 
 while its dynamics is qualitatively well described  
by an analytical method, it cannot reproduce correctly the libration frequency. Instead, our model correctly predicts the libration period providing a good complement to the analytical theory.
 We will show some applications of our model in the next sections.

\section{Application to planetary systems}\label{sec.planets}

The resonant model described in this paper has several different 
applications. While it may be employed to study isolated commensurabilities, 
giving information about the equilibrium solutions, libration periods and 
widths, it may also be used to map a region of the phase space. It may
therefore aid in quantifying the proximity of the system to resonant
configurations and identify regions susceptible to instabilities stemming
from resonance overlap. In certain ways, the model allows to obtain a
qualitative description of the resonant structure of the phase plane for a
fraction of the computational cost of a dynamical map \citep[e.g. Megno, Mean Exponential Growth of Nearby Orbits,][]{Cincotta.Simo.1999}.

\subsection{The HD31527 system}
\label{subsec:hd3}

In our first application we analyze the orbital evolution and resonant 
structure in the vicinity of the HD31527 system. Discovered by \cite{Mayor+2011}
and recently revisited by \cite{Coffinet+2019}, this star hosts three 
Neptune-size planets orbiting a G0V star with mass $m_0 = 0.96 M_\odot$. Even
though these publications do not give all the planetary parameters, it is 
possible to obtain a best-fit of the radial velocity data thanks to the Data
\& Analysis Center for Exoplanets\footnote{\url{https://dace.unige.ch}} 
\citep[DACE; see][]{Buchschacher+2015}. Keplerian model initial conditions
were computed using the formalism described in \cite{Delisle+2016}. We added 
quadratically a noise level of $0.75$ m/s to the published radial velocities (RV) uncertainties
to take into account stellar jitter and the typical instrumental noise floor
seen in HARPS data. In addition to the signals fitted, we allowed an RV
offset. Night binning reduces from 256 to 244 the RV data, although orbital
fit remained virtually unchanged.

Results are shown in Table \ref{tab:hd31527}, where the error-bars in
semimajor axes and masses were estimated from the joint posterior
distribution of the model parameters using the Markov chain Monte Carlo (MCMC) analysis \citep{Diaz.2014}.
Compared with previous estimates, the greatest change in the orbital fit lies
in the eccentricity of the outer planet. \cite{Mayor+2011} found $e_d \simeq
0.38$ while \cite{Coffinet+2019} gave $e_d \simeq 0.67$. Since neither of
these works presented estimates for mean longitudes or longitudes of
pericenter, no stability check was carried out. In fact, and to the best of
our knowledge, no dynamical studies have been performed and its long-term
orbital evolution has yet to be determined. Following usual conventions, for
our dynamical analysis we will denote the inner planet (HD31527 b) as $m_1$,
the middle body (HD31527 c) by $m_2$ and the outermost (HD31527 d) by $m_3$. 

\begin{table}
\centering
\caption{Minimum masses and orbits for the HD31527 system. The mass of the star is $m_0 = 0.96 M_\odot$.}
\label{tab:hd31527}
\resizebox{0.49\textwidth}{!}{
\begin{tabular}{|c|c|c|c|c}
\hline
parameter   &      HD31527 b       &       HD31527 c      &      HD31527 d    \\
\hline
 $K$  [m/s] &   2.818 $\pm$ 0.092  &   2.661 $\pm$ 0.093  &  1.70 $\pm$ 0.20  \\
 $P$  [d]   & 16.5545 $\pm$ 0.0024 &  51.265 $\pm$ 0.023  & 272.84 $\pm$ 0.78 \\
 $e$         &  0.137 $\pm$ 0.033  &   0.030 $\pm$ 0.034  &  0.596 $\pm$ 0.055\\
 $\omega$ [deg] & 47 $\pm$ 14      &     277 $\pm$ 69     &  183.3 $\pm$ 6.5  \\
 $T_P$  [d]  &      55499.7379     &       55466.7627     &    55444.7471     \\
\hline
 $m \sin{I}$ [$m_\oplus$] & 10.8279 $\pm$ 0.50 & 15.0399 $\pm$ 0.71  &   
13.5044 
$\pm$ 1.20       \\
$a$ [au]     &  0.1254 $\pm$ 0.002 & 0.2664 $\pm$ 0.005  & 0.8121 $\pm$ 0.014 \\
\hline
planet \#     &             1       &        2             &    3 \\
\hline
\end{tabular}}
\end{table}

\begin{figure}[ht!]
\centering
\includegraphics[width=0.99\columnwidth,clip]{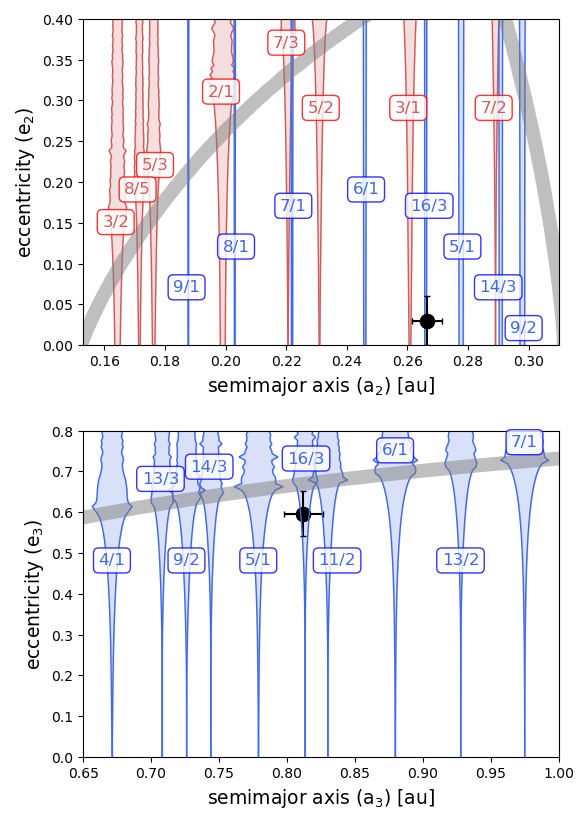}
\caption{Resonant structure determined by the semi-analytical model in the 
$(a_2,e_2)$ plane surrounding the middle planet (top) and in the $(a_3,e_3)$ 
plane (bottom). Two-planet commensurabilities between planets $m_1$ and $m_2$
are shown in red while those between the middle and outer planets are
presented in blue. In both plots the nominal position of the planets are
identified by black filled circles, while the collision curves are depicted
in gray.}
\label{fig.hd31527.1}
\end{figure}

The top graph of Fig. \ref{fig.hd31527.1} shows the resonant structure in
the $(a_2,e_2)$ plane around the nominal position of the middle planet. The 
libration width of each commensurability was determined from Eqs. 
(\ref{dela1ja})-(\ref{dela2ja}) assuming constant semimajor axes, 
eccentricities and longitudes of pericenter. Mutual inclinations were taken 
equal to zero and all orbital elements (with the exception of $a_2$ and
$e_2$) were assumed equal to their nominal values. Since the resonance widths
are determined by the maximum range attained by the orbit-averaged disturbing
function (i.e. $\Delta a_i \propto \sqrt{\Delta R} = \sqrt{R_{\rm max} - R_{\rm min}}$),
the libration width increase for initial conditions closer to the collision
curves, highlighted in gray. For eccentricities beyond this curve, the
maximum value of $R$ can attain very large values that may not be correctly
estimated using finite differences. This leads to a resonant structure that
appears jagged and irregular and unreliable. Moreover, very large values of
$R$ are associated with close encounters between the planets. and
instabilities. In the absence of a suitable protective mechanism (i.e.
mean-motion resonance), all trajectories with $e_2$ above the collision curves
would lead to a disruption of the system. 

It is important to mention that throughout this work the collision curve
was determined numerically from the calculation of the non-averaged disturbing
function. For each value of the semimajor axis, we searched for the minimum
eccentricity for which the mutual distance between the bodies was zero for some
combination of both mean longitudes. Only within the libration region of a MMR,
is it possible that the correlation between the mean longitudes allow the
planets to avoid the collision point, in a manner similar as Neptune and Pluto
are able to evade close encounters due to their resonant configuration.

Mean-motion resonances (MMR) with the inner planet are colored red and their 
width can be seen to decrease towards the right-hand side of the plot. 
Conversely, mean-motion commensurabilities with the outer planet are shown in
blue. Significantly thinner due to their higher order, these MMR become less 
relevant for smaller values of the semimajor axis $a_2$. 

Due to the relatively small planetary masses, most resonances are thin and
the phase space below the collision curve appears dominated by secular
interactions. Even the 3/1 MMR between $m_1$ and $m_2$ is barely visible for
eccentricities of the order of $e_2$, and its effects are therefore not
expected to be significant. Curiously, the nominal location of $m_2$
practically coincides with the 16/3 mean-motion commensurability with the
outer planet. Again, however, the width of the MMR in the plot is very small,
and there does not appear to be much indication of any significant effect on
the dynamics of the system.

The bottom graph of Fig. \ref{fig.hd31527.1} shows a similar resonant map, 
but now for the semimajor axis and eccentricity plane of the outer body. Only
MMRs with $m_2$ are plotted. Even though all commensurabilities are of a high
order, the large eccentricity of the planet increases their reach, generating
a significant widening of the libration region for values of $e_3$ close to
the nominal value. The best-fit solution lies close to the collision curve,
with the top of the error-bar almost reaching this limit. As in the previous
frame, the nominal semimajor axis almost coincides with that of the 16/3 MMR,
whose width in this plane is substantially larger than previously seen in
$(a_2,e_2)$. Perhaps some influence in the dynamical evolution of the system
is possible after all.

\begin{figure}[ht!]
\centering
\includegraphics[width=1.\columnwidth,clip]{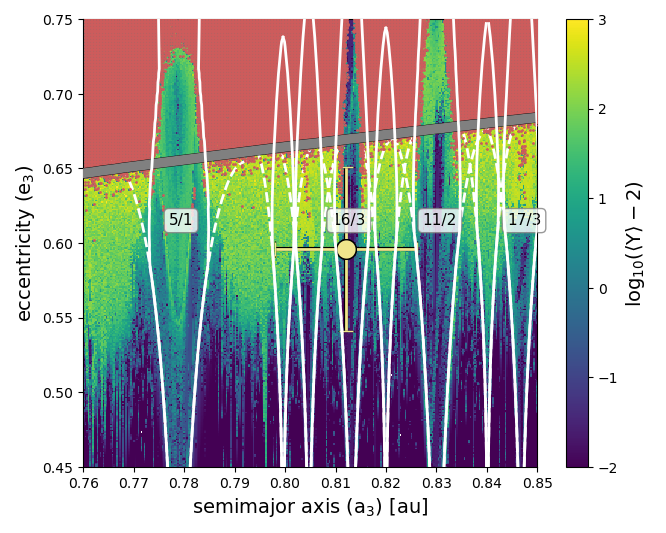}
\caption{Megno map for the HD31527 system, as obtained from the numerical 
integration of a grid of $300 \times 300$ initial conditions in the
$(a_3,e_3)$ plane. All other orbital elements were set to the nominal values.
Integration time was set to $10^4$ years, and escapes during this timespan
are highlighted in red. The color code shows values of $(\langle Y \rangle -
2)$. Dark blue colors correspond to more regular orbits while lighter tones
of green indicate increasingly chaotic motion. The location of the outer
planet is shown with a yellow filled circle and the collision curve in gray.
The libration widths of the most relevant MMRs determined analytically are
plotted in white lines. See text for details.}
\label{fig.hd31527.2}
\end{figure}

Figure \ref{fig.hd31527.2} shows a Megno \citep{Cincotta.Simo.1999} dynamical
map, again in the $(a_3,e_3)$ plane, but now focusing on a smaller region 
around the outer planet. A total of $300 \times 300$ initial conditions were 
integrated for $10^4$ years. All other orbital elements, including the
initial mean anomalies, were set to those given in Table \ref{tab:hd31527}.
Initial conditions leading to escapes/collisions during the simulation are
colored red. Regular motion (at least within the integration span) is shown
in dark colors, while increasingly chaotic trajectories identified with
lighter tones. Note how only the central region of the stronger MMRs prove a
protective mechanism sufficiently effective to avoid disruptive perturbations
and close encounters. These safe zones appear as green stalagmites above the
collision curve, shown in gray. 

\begin{figure*}[ht!]
\centering
\includegraphics[width=1.98\columnwidth,clip]{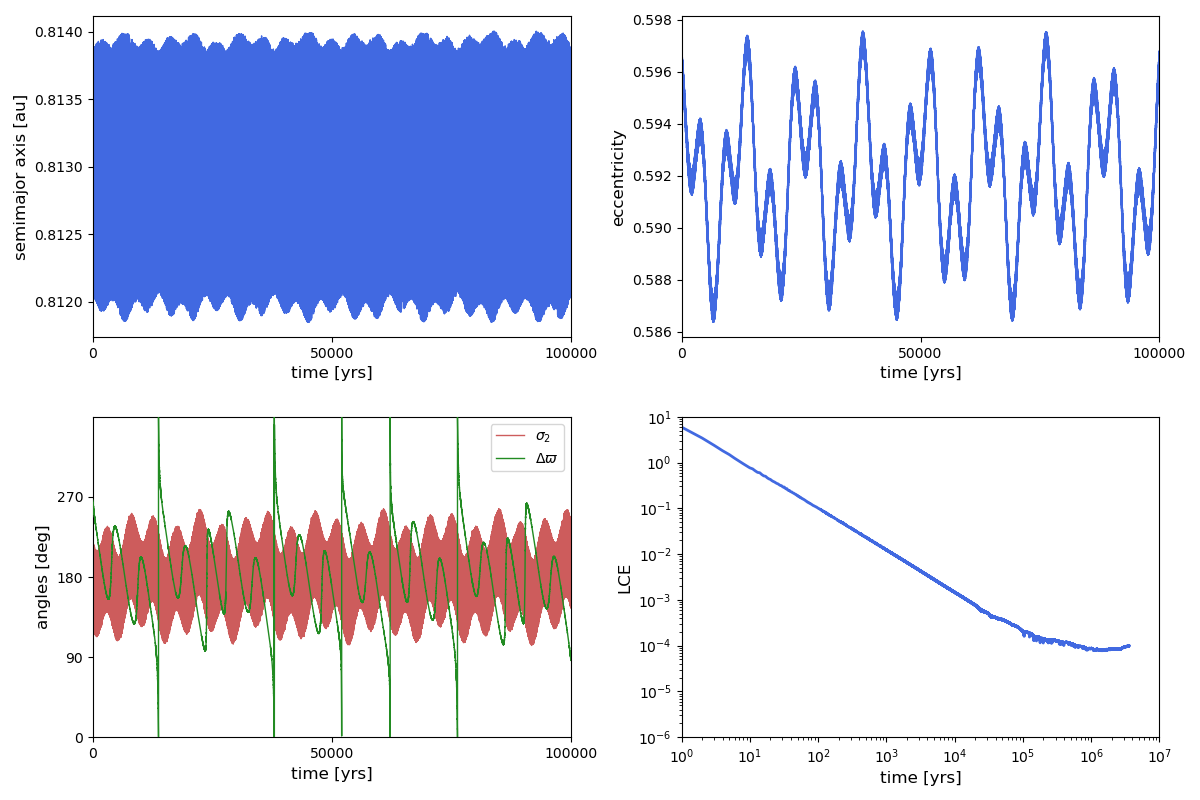}
\caption{Results of a long-term N-body simulation of the nominal solution for
the HD31527 system. Top rows show the semimajor axis $a_3$ and eccentricity 
$e_3$ of the outer planet, as a function of time, for the first $10^5$ years
of the integration. The lower left-hand plot shows, in the green, the
difference in longitude of pericenter $\varpi_3-\varpi_2$, while the red
curve presents the behavior of the resonant angle $\sigma_2 = 3 \lambda_2 - 
16 \lambda_3 + 13\varpi_3$. The lower right-hand graph shows the maximum
Lyapunov exponent (LCE) for the system during the complete simulation.}
\label{fig.hd31527.3}
\end{figure*}

The separatrix of several mean-motion resonances are shown in white lines, 
the most important of which are identified in rounded labels. For the 
estimation of the libration widths we employed two different approximations. 
Dashed lines show results obtained without specifying any upper bounds in 
$R_{\rm max}$, therefore yielding results which are analogous to those in
Fig. \ref{fig.hd31527.1}. As before, the resonance widths increase
monotonically for higher eccentricities, leading to an overlap of adjacent
MMRs for initial conditions close to the collision curve. Since the
structures beyond this point may not be reliable, they are not plotted. 

In a second calculation, the libration widths were calculated limiting the 
value of $R_{\rm max}$ to configurations where the distance between the
planets was larger than a lower bound $\Delta_{\rm min}$. With this approach
we aimed to reduce the reach of the commensurability to include only those
initial conditions that avoid close encounters and, thus, indicative of
stable motion.

Without a detailed stability criteria, the necessary value of $\Delta_{\rm 
min}$ can only be estimated qualitatively. For the restricted 3-body problem,
 \cite{2020CeMDA.132....9G}  showed that good agreements with
N-body simulations were found using $\Delta_{\rm min} \sim 3 R_{\rm Hill}$,
where $R_{\rm Hill}$ is the Hill radius of the perturber. Following the same
idea, in the planetary case involving bodies $m_i$ and $m_j$, we will adopt 
$\Delta_{\rm min} = 2 \sqrt{3} R^{(i,j)}_{\rm Hill}$, where
\be
R_{\rm Hill}^{(i,j)} = \frac{a_i + a_j}{2} \biggl[ \frac{m_i+m_j}{3m_0} 
\biggr]^{1/3}
\ee
is the so-called mutual Hill radius of the planets 
\citep{Marzari.Weidenschilling.2002}. This expression for $\Delta_{\rm min}$
is equal to the minimum distance between planets in circular orbits leading
to Hill-stability, as deduced originally by \cite{Marchal.Saari.1975} and 
\cite{Marchal.Bozis.1982}, and later popularized by \cite{Gladman.1993}.
Although this choice for $\Delta_{\rm min}$ does not constitute a credible
stability limit for multi-planet systems in eccentric orbits, we found it a
simple expression leading to reasonable agreements with N-body simulations. A
more rigurous analysis is left for future studies.    

The libration widths incorporating $\Delta_{\rm min}$ as a minimum distance 
between the planets are shown with continuous white lines in Fig. 
\ref{fig.hd31527.2}. The effect of the cut-off starts being noticeable well 
below the collision curve and leads to a reduction of the effective size 
of the stable resonance domain. For the 5/1, 16/3 and 11/2 MMRs the
analytical results show an excellent agreement with the dynamical map, even
if the predicted regions are overestimated. The modified model also predicts
finite libration widths for higher-order and weaker commensurabilities;
however no corresponding structure is observed in the N-body map. 

The nominal position of the outer planet is shown with a large yellow filled 
circle. The error-bars are very evident in both the semimajor axis and the 
eccentricity. The best-fit coincides almost exactly with the 16/3 MMR, but 
non-resonant trajectories also lie within the error bars. However, most of
these solutions are associated to highly chaotic motion while only the 16/3
MMR appears regular. As shown by the white lines, the regions leading to
chaotic motion are defined by high-order commensurabilities. The correlation
in the libration widths between the dynamical map and the semi-analytical model
is very good, and proves a very useful tool to map the overall dynamical
characteristics of the system. 

To study the resonant motion and long-term stability of the system, we 
performed an N-body simulation of the nominal solution for $\sim 3 \times
10^6$ years. The calculation of the maximum Lyapunov exponent (LCE) is shown
in the lower right-hand frame of Fig. \ref{fig.hd31527.3}, yielding a
characteristic time of the order of $\tau_{\rm LCE} \sim 10^4$ years. This
indicates that the best-fit system is only weakly chaotic and the departure
from a purely regular trajectory should only be noticeable for timescales
larger that $\tau_{\rm LCE}$, in agreement with the estimated value of
$\langle Y \rangle$ shown in the previous figure.

The time evolution of the semimajor axis and eccentricity of the outer planet
are shown in the top row of Figure \ref{fig.hd31527.3}; in order to aid the 
visualization of the resonant and secular periods, only the first $10^5$
years of the integration are plotted. The solution appears completely
quasi-periodic and we found no trace of a secular trend or long-term increase
in the amplitudes. Finally, the bottom left-hand graph shows, in green, the
behavior of the secular angle $\Delta \varpi_{(3,2)} = \varpi_3 - \varpi_2$
while the red curve corresponds to the resonant angle
\be
\sigma_2 = 3 \lambda_2 - 16 \lambda_3 + 13\varpi_3 .
\ee

From these results, it appears that the middle and outer planets of the
HD31527 system lie inside the 16/3 MMR displaying a moderate-amplitude
libration around $\sigma_2 = \pi$. To the best of our knowledge, this is the
first time an exoplanetary system is found in such a high-order and
high-degree mean-motion commensurability, and whose resonant motion functions
as an effective protection mechanism. The chaoticity of the surrounding
regions are so strong that it is difficult to imagine a formation scenario
for this system involving planetary migration, even if this process was
responsible for the excitation of the eccentricity. Two stronger MMR (5/1 and
11/2) are found close by and to both sides of the present-day position; it is
strange to imagine how these could have been avoided while capturing the
system in the (weaker) 16/3 resonance. 

The difference in longitude of pericenter between these planets shows an 
oscillation around $\Delta \varpi_{(3,2)} = \pi$ with a secular period of the
order of $T_{\rm sec} \simeq 5000$ years. When the eccentricity of the middle
planet approaches zero, the secular angle shows a spike and a rapid
circulation. This behavior, however, is not related to any chaotic motion.

The libration period may also be estimated with the semi-analytical model and
yields a value of the order of $T_{\rm lib} \simeq 22$ years, while a Fourier
decomposition of the N-body simulation yields a value close to $\sim 19$
years. We believe the agreement is very satisfactory, particularly
considering that the analytical estimate corresponds to zero-amplitude
librations while the numerical solution has a considerable oscillation around
the fixed point. We believe this calculation shows additional evidence on how
the model may be used to obtain qualitative information on the resonant
dynamics, whatever the eccentricity or mutual inclination of the bodies.

\subsection{The HD74156 system}
\label{subsec:hd7}

In the early 2000s it was suggested that most planetary systems should be 
dynamically packed \citep[e.g.,][]{Barnes.Quinn.2001,Barnes.Raymond.2004},
with little room for additional bodies between adjacent planets \citep[e.g.,][]{Giuppone+2013}. Based on
this idea, several works analyzed the phase space of known systems searching
for stable regions that may house undetected planets. One of these systems
was HD74156. 

Discovered by \cite{Naef.etal.2004}, it currently has two confirmed planets 
orbiting a $m_0 = 1.24 M_\odot$ star. One of the bodies lies close to the star 
with an orbital period of the order of $P_b \simeq 60$ days while the second is 
found significantly further out ($P_c \simeq 5.5$ years). After a series of 
N-body simulations, \cite{Raymond.Barnes.2005} found that a Saturn-type planet 
could well exist between both known bodies and the resulting system would be 
stable for timescales of the order of the age of the star. 

As new observations became available, \cite{Barnes.etal.2008} extended the 
numerical study and further constrained the mass and orbit of the
hypothetical third planet to the new radial velocity data. They found that
the region between the two confirmed planets contains a rich dynamical
structure defined by both isolated and interacting mean-motion resonances
with both bodies. The most promising system configuration is reproduced in Table \ref{tab:hd74156}.

\begin{table}
\centering
\caption{Masses and orbital parameters of the 3-planet HD74156 system proposed 
by \citep{Barnes.etal.2008}. The mass of the star is $m_0 = 1.24 M_\odot$.}
\label{tab:hd74156}
\begin{tabular}{|c|c|c|c|c}
\hline
   parameter          &  HD74156 b  &  HD74156 d  &  HD74156 c  \\
\hline
 $m$ [$m_{\rm Jup}$]  &     1.847   &    0.412    &    7.995    \\
    $a$ [au]          &     0.292   &    1.023    &    3.848    \\
    $e$               &     0.629   &    0.227    &    0.426    \\
 $\omega$ [deg]       &   176.45    &  191.81     &  262.17     \\
   $M$  [deg]         &   211.64    &   67.51     &  340.20     \\ 
\hline
planet \#     &             1       &        2             &    3 \\
\hline
\end{tabular}
\end{table}

Unfortunately, subsequent orbital fits including stellar jitter 
\citep{Baluev.2009} did not confirm planet "d" arguing that its signal could
be due to annual systematic errors from High Resolution Spectrograph
(HRS) of the Hobby–Eberly Telescope. \citet{Wittenmyer.2009}, \citet{ Meschiari.etal.2011}, and \citet{Feng.2015} later updated the system with additional observations and also failed to detect the proposed middle planet. 

\begin{figure*}[ht!]
\centering
\includegraphics[width=2\columnwidth,clip]{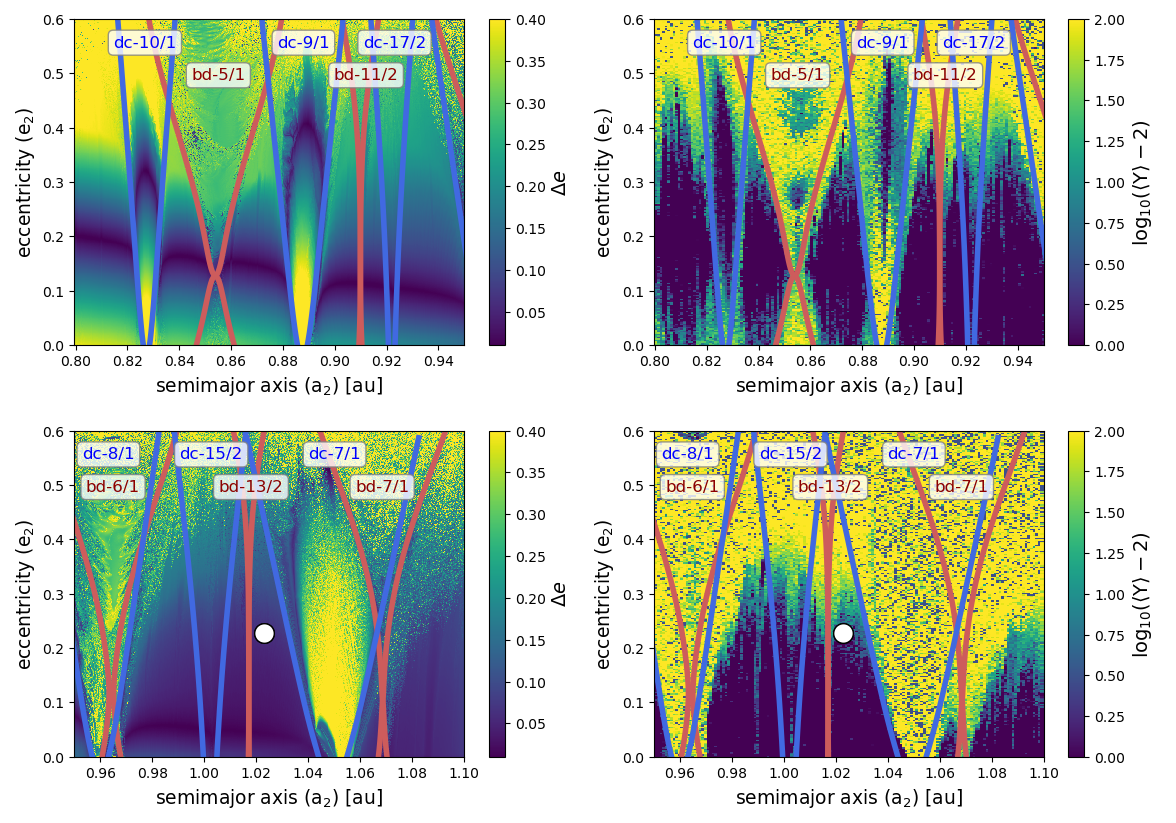}
\caption{ Left: Top and bottom graphs show a $\Delta e$ dynamical map for
the HD74156 system, as obtained from the numerical integration of a grid of $900
\times 900$ initial conditions in the $(a_2,e_2)$ plane. Upper plot corresponds
to semimajor axis $a_2 \in [0.80,0.95]$ au, while the bottom frame for $a_2 \in
[0.95,1.10]$ au. Integration time was set to $10^3$ years, and escapes during
this timespan are highlighted in light yellow. The location of the middle planet
is shown with a white filled circle. The libration widths of the most relevant
MMRs determined analytically are highlighted in blue (resonances with outer
planet) and red (resonance with inner planet). Right: Megno map for 
the same plane. The set of initial conditions was reduced to a $300 \times 300$ grid, 
but the integration time extended to $10^4$ yrs.}
\label{fig.hd74156.1}
\end{figure*}

However, the resulting dynamical structure of the 3-planet system proposed by
\cite{Barnes.etal.2008} proved sufficiently interesting to be a good test for
our  model. Even if the third planet may not actually exist, the 
hypothetical extended system constitutes a challenging benchmark for the
predictive power of the model, as well as its capacity of identifying
underlying features in the dynamical maps.

As before, we will number the planets according to their distance to the 
central star, as identified in the lower row of Table \ref{tab:hd74156}. The 
confirmed planets are thus $m_1$ and $m_3$ while the proposed unconfirmed body
is labeled $m_2$. Since we will study a hypothetical system, we need not be
tied to specific values for the angles. We will therefore modify the orbital
elements given in Table \ref{tab:hd74156} such that
\be
M_1 = M_2 = M_3 = 0 \hspace*{0.3cm} ; \hspace*{0.3cm}  \varpi_1 = \varpi_2 = 0 
                    \hspace*{0.3cm} ; \hspace*{0.3cm} \varpi_3 = \pi .
\ee
In other words, all planets are assumed to lie on their pericenters and the 
difference in the initial apsidal orientation between adjacent bodies is
taken to be: $\Delta \varpi_{(2,1)} = \varpi_2 - \varpi_1 = 0$ and $\Delta 
\varpi_{(3,2)} = \varpi_3 - \varpi_2 = \pi$. This choice helps preserve the 
symmetry in the libration regions in the dynamical map and, in some cases, 
maximizes their width. 

For the representative plane, we chose the $(a_2,e_2)$ plane, more or less 
centered around the proposed position of the middle planet, and with limits 
similar to those used in Fig. 2 of \cite{Barnes.etal.2008}. The left-hand
plots of Fig. \ref{fig.hd74156.1} show color plots of $\Delta e$ for a grid
of $900 \times 900$ initial conditions varying the values of the semimajor
axis $a_2$ and eccentricity $e_2$ while maintaining all other orbital elements
fixed. Each initial condition was then integrated for $10^3$ years, keeping
track of any ejections or collisions during this time. The color code 
corresponds to the maximum change in eccentricity of $m_2$ during the
integration (i.e. $\Delta e = {\rm max}(e_2) - {\rm min}(e_2)$), and has
proven to be a powerful indicator of secular and resonant dynamics
\citep[e.g.][]{Dvorak.etal.2004,Ramos.etal.2015}. 

While the vertical light-tones plumes are generated by the action of
mean-motion resonances, the dark colored diagonal stripe in the upper 
 left-hand plot represents the so-called Mode I secular mode 
\citep{Michtchenko.Malhotra.2004} between the inner and middle planet
and represents the locus of initial conditions leading to zero-amplitude 
oscillation of the secular angle $\Delta \varpi_{(2,1)}$ around zero. 
In the plot the Mode I secular mode may be seen as a diagonal curve running 
from $(a_2,e_2) \simeq (0.8,0.2)$ down to $(a_2,e_2) \simeq (0.95,0.1)$.
An analogous, but less evident, stripe may also be seen in the lower 
left-hand plot for larger values of the semimajor axis. Since the 
longitude of pericenter of the outer planet was chosen such that 
$\Delta \varpi_{(3,2)} = \pi$, no indication of a Mode I with the outer 
planet is observed.

For the right-hand plots, we reduced the grid to a set of $300 \times 300$
initial conditions but extended the integration time to $10^4$ years. The
color code now shows the MEGNO values at the end of the simulation. As in the
dynamical map constructed for the previous planetary system, dark blue colors
correspond to regular orbits, while lighter tones represent increasingly
chaotic trajectories. In both set of graphs we the semimajor axis domain was
split in two sections: the region with $a_2 \le 0.95$ au is shown on the top
frames, while the bottom graphs show results for $a_2 \ge 0.95$ au. The
position of the proposed planet is highlighted with white filled circles.

The maximum widths of the relevant MMRs, as obtained with the 
model, are shown in red lines (for commensurabilities with the inner planet)
and blue lines (for resonances with the outer planet). These are identified
on the top part of the graphs using text with the corresponding color. In
most cases we note a very good agreement between the predicted resonant
structure and the dynamical maps. For example, the hour-glass figure centered at
$a_2 \simeq 0.86$ au caused by the 5/1 MMR with the inner planet. For
eccentricities $e_2 \lesssim 0.13$, the stable equilibrium solution occurs
for $\sigma = \lambda_1 - 5 \lambda_2 + 4\varpi_2 = \pi$, resulting in
chaotic trajectories inside the separatrix. Conversely, for higher
eccentricities the stable solution occurs for values of the resonant angle
$\sigma=0$ and thus coincidental with the angles used to define the
representative plane. The central region of the resonance is therefore more
regular and admits librational solutions. 

These "hour-glass" structure observed for the 5/1 MMR are not new and were
detected in studies of circumbinary planets by \cite{Zoppetti.etal.2018}. The
center of the "X" shape appears to coincide with the forced eccentricity of the
planet \citep{Zoppetti.et.al.2019}, a correlation that also appears in the plots shown here.

Another interesting feature is the large chaotic region located at $a_2
\simeq 0.96$ au and visible in the lower plot. The resonance analysis shows
that it is generated by the interaction between two distinct
commensurabilities, the 6/1 between planets $m_1$ and $m_2$, ad the 8/1
between $m_2$ and $m_3$. A similar structure occurs at $a_2 \simeq 1.06$ au
between two 7/1 MMRs, only this time the overlap does not occur for
near-circular orbits, leading to a wedge-shape region of regular motion at
low eccentricities.

Moving to a more general analysis of the phase plane, recall that in a $n_i/n_j = k_2/k_1$ MMR there are two independent resonant angles, which we may denote by:
\be
\begin{split}
\sigma_1 =& k_1 \lambda_i - k_2 \lambda_j + (k_2-k_1) \varpi_i \\
\sigma_2 =& k_1 \lambda_i - k_2 \lambda_j + (k_2-k_1) \varpi_j .
\end{split}
\ee
All other possible critical angles may be written as linear combinations of
$\sigma_1$, $\sigma_2$ and the difference in longitudes of pericenter. 
Our model analyzes the behavior of the resonances based on the study of the
combination $k_1 \lambda_i - k_2 \lambda_j$, with independence of the
definition of $\sigma$. Anyway, we can calculate the location of the
equilibrium points corresponding to $\sigma_1$ for example and we can  analyze
how the predictions of this model correlate with the dynamical structures
found in the Megno map. Nevertheless, the  model can also be used to
separate the mean-motion resonances into three distinct groups, according to
the number of stable solutions of $\sigma = \sigma_1$ and their equilibrium
values. These are detailed below.

\begin{itemize}

\item \underline{$\sigma = 0$ stable solutions}

These are 2-planet commensurabilities, between the middle planet $m_2$ and
one of its neighbors, in which a the stable equilibrium solution corresponds
to a value of the resonant angle $\sigma=0$. Examples are: dc-10/1, bd-11/2, 
dc-17/2, bd-6/1, dc-15/2 and bd-13/2. Following the notation used in the
figure, the letters indicate the planets involved in the MMR. These solutions
are found to be stable for all the values of $e_2$ up to the edge of the
collision curve. In some cases, these solutions are stable only for a certain
range of eccentricities. These include bd-5/1 (for $e_2 \gtrsim 0.13$),
dc-8/1 ($e_2 \gtrsim 0.01$) and bd-7/1 ($e_2 \gtrsim 0.1$).
 
\vspace*{0.2cm}
\item \underline{$\sigma = \pi$ stable solutions}

Similar to the previous case, these resonances are characterized by a single 
stable solution located at $\sigma=\pi$. An example is dc-9/1. Most of these 
solutions, however, are found for a limited interval of the eccentricity
$e_2$. These include bd-5/1 (for $e_2 \lesssim 0.13$), bd-7/1 ($e_2 \lesssim
0.1$), dc-7/1 ($e_2 \gtrsim 0.05$). In these cases the solutions found for
other eccentricities usually correspond to $\sigma=0$ librations.

\vspace*{0.2cm}
\item \underline{Asymmetric stable solutions}

This group harbors fixed points having values of $\sigma$ different from $0$
or $\pi$. Examples were found only in resonances with the outer planet and 
include dc-7/1 (for $e_2 \lesssim 0.05$) and dc-8/1 (for $e_2 \lesssim
0.01$). For higher eccentricities these solutions reduce to symmetric
librations, either with $\sigma=0$ or $\sigma=\pi$.
 
\end{itemize}

In general the behavior of resonant initial conditions in the dynamical map 
show a very good agreement those expected from the model. Thus, resonances
with stable solutions in $\sigma=k_2 \pi$ are expected to show regular solution
inside the libration domain for the chosen representative plane. Such behavior is 
clearly observed in both the dc-10/1 and dc-9/1 commensurabilities and, in lesser
degree, in the bd-5/1 resonance. 

Conversely, resonances with stable solutions in $\sigma=\pi$ should exhibit 
more chaotic motion in the representative plane. This predictions is observed
in the case of dc-7/1 and, for example, low eccentricity initial conditions
in the bd-5/1 resonance. It is important to recall, however, that there is no unique
resonant angle, particularly for high-order MMRs, and the most dominant combination
between the mean longitudes and longitudes of pericenter depends on the
eccentricities and relative masses of the planets involved in the commensurability.
Even so, the information that the model yields still serves as an initial analysis of
the resonance, and serves to identify characteristics or interesting features that
deserve further study.

\section{Application to circumbinary planets}\label{sec.cbp}
The Kepler mission revolutionized our knowledge about the existence of circumbinary planets (CBPs), with the discovery of 13 transiting CBPs orbiting 11 Kepler eclipsing binaries. Recently the first circumbinary planet was reported using the data from TESS mission \citep[TOI~1338,][]{Kostov+2020}. These planets brings new challenges on formation theories and their formation mechanism is still under debate, because most of them lie near their instability boundary at about 3 to 5 binary separations. Recent studies argue that the occurrence rate of giant, Kepler-like CBPs is comparable to that of giant planets in single-star systems \citep[see][and references therein]{Martin+2019}.

From this sample, all the transiting binaries that host planets have short periods ($<30$ days) \citep{Schwarz+2016} with eccentricities ranging from quasi circular orbits (like \textit{Kepler}-47, \citealt{Orosz+2012}) to highly eccentric (like \textit{Kepler}-34, \citealt{Welsh+2012}). Typical planets detected around binary stellar systems have a radius of the order of $10$ Earth radii and orbital periods of about $160$ days, on almost circular coplanar orbits (i.e. $a_{\rm p} \sim 0.35$ au), ~\citet{Martin2018}. 

Transiting inclined CBPs are more difficult to detect because the bias selection. However, planets do form in circumbinary discs (CBDs) and observations of highly non-coplanar CBDs, such as 99~Herculis \citep{Kennedy+2012a}, IRS~43 \citep{Brinch+2016}, GG~Tau \citep{Cazzoletti+2017, Aly+2018}, HD~142527 \citep{Avenhaus+2017}, and HD~98800 \citep{Kennedy+2019} were detected. 

Former analysis of the stability of inclined \textit{massless} particles was done by~\cite{Doolin+2011}, where they considered different binary eccentricities and mass ratios between its components. More recently \cite{Cuello&Giuppone2019} and \cite{Giuppone+Cuello.2019} studied the evolution of CBDs and the stability of inclined Jupiter-like planets around binaries. Recently, circumbinary retrograde coplanar stability was addressed by \citet{Hong+2021} for circular and eccentric planets. Furthermore, there is some evidence that the $N/1$ MMRs are related with planetary parking location and their evolution in the disk. With a simplified model \citet{Zoppetti.etal.2018} suggested capture in 5/1 MMR and subsequent tidal evolution of the planet, whereas some effort with fine-tuning hydrodynamical simulations tried to put some constrains in the protoplanetary disks that allows migration of CB planets, but not showing capture at MMRs \citep[see e.g.,][and references therein]{Penzlin+2020}.

In the present section we study the $N/1$ and $N/2$ mean motion resonances (identified as $k_2/k_1$ in the model), and our aim is to correlate unstable regions near the binary with the MMR overlap. 

The resonant model described in this paper has several different 
applications. While it may be employed to study isolated commensurabilities, giving information about the equilibrium solutions, libration periods and 
widths, it may also be used to map a region of the phase space. It may
therefore aid in quantifying the proximity of the system to resonant
configurations and identify regions susceptible to instabilities stemming
from resonance overlap. In certain ways, the model allows to obtain a
qualitative description of the resonant structure of the phase plane for a
fraction of the computational cost of a dynamical (e.g. Megno) map.

We use the model described in section \ref{jacobi} to map a region of the phase space, identifying regions susceptible to instabilities stemming
from resonance overlap. Regular motion and chaos of a coplanar CBP in the plane $(\alpha,e)$ are presented in Section \ref{sec:cbp-coplanar} (with $\alpha=a/a_B$, whereas we studied inclined CBP in the plane $(a,i)$ initially placed in circular orbit in Section \ref{sec:cbp-inclined}. We consider different kind of binaries such those detected by Kepler/TESS. We do not expect effects on the binary orbit, because the mass of the planet is much lower than the binary companion. To describe the orbit of the binary we use orbital elements with sub-index $B$, and orbital elements without sub-index for the planet. Without loss of generality, we set initial conditions \{$\lambda_{\rm B}=0\degr$, $\varpi_{\rm B}=0\degr$, $\Omega_{\rm B}=0\degr$\}. Our fiducial binary has $a_B=0.1$ au, and the sum of their masses equal to 1. The angles of a given planet are measured from the direction of the binary pericentre and Jacobi orbital elements of circumbinary planet given without subscripts. 

\subsection{Coplanar circumbinary planets}\label{sec:cbp-coplanar}

\begin{figure}[ht]
\centering
\includegraphics[width=0.98\columnwidth,clip]{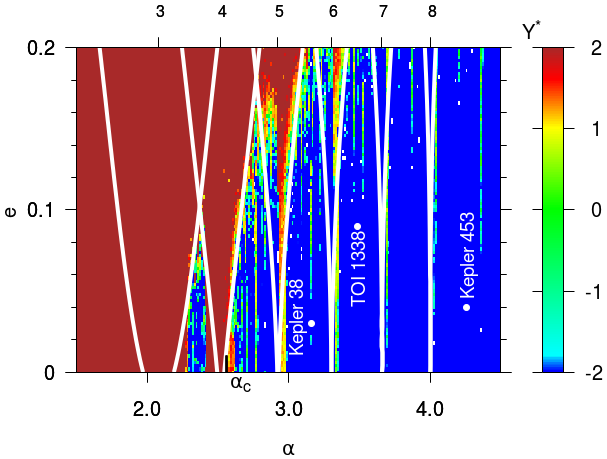}
\caption{Megno map for circumbinary system, as obtained from the numerical integration of a grid of $250 \times 100$ initial conditions in the $(a,e)$ plane. All angles are initially set to zero. Integration time was set to $10^4$ binary periods, and escapes during this timespan are highlighted in brown. The color code shows values of $Y^*=\text{log}_{10}(\langle Y \rangle-2)$. Dark-blue colors correspond to more regular orbits while lighter tones of green indicate increasingly chaotic motion. The libration widths of the most relevant MMRs determined with the model are plotted in white lines. Top scale indicates the positions of $N/1$ resonances with respect of the mean motion of the binary. See text for details.}
\label{fig.CBplanets}
\end{figure}

 In this subsection we start with coplanar orbits and we only show the location of MMR in the plane $(\alpha,e)$ for a binary with mass ratio $q=0.2$ and eccentricity $e_B=0.1$, being $\alpha=a/a_B$. Figure \ref{fig.CBplanets} shows a Megno dynamical map for a planet, focusing on a region between $1.5 < \alpha < 4.5$. A total of $250 \times 100$ initial conditions were integrated for $10^4$ binary periods. All other angular elements were set to zero. Again, initial conditions leading to escapes/collisions during the simulation are colored brown. Regular motion (at least within the integration span) is shown in blue colors, while increasingly chaotic trajectories identified with green/yellow tones. We also overlay with white lines the N/1 resonances with $N>3$ setting $\varpi=180^o$. 

The semi-analytical model predicts that all the $N/1$ MMRs have one stable center of resonance around $0^\circ$, in the region plotted in Fig. \ref{fig.CBplanets}. However, maybe because the fast precession of $\varpi$ induced by the binary \citep[see][]{Farago&Laskar2010, Li+2014}, closer resonances are unstable ($3/1$ and $4/1$). Contrary to planetary case, we identify that regions inside of the stronger $N/1$ MMRs are chaotic (mainly observed in MMR $3/1$, $4/1$). Overlap of $3/1$, $4/1$, and $5/1$ delimits the unstable regions in the plane ($\alpha,e$) showing a perfect agreement of the model with numerical integrations. The $V$-shape of each resonances is evident and resonant-width decreases as we increase $N$, being almost zero for circular planets. 

Additionally, in the Figure \ref{fig.CBplanets} we plot the position of some similar CB systems with orbital parameters resumed in Table \ref{tab:CB}. We indicate with $\alpha_c$ the minimum stable coplanar-prograde semimajor axis, using the criteria of \citet{Holman+Wiegert1999}. This empirical criteria, widely used, is based in a polynomial fit that considers quadratical exponents of $a_B$, $e_B$ and $q$. According to our Figure \ref{fig.CBplanets} this limit roughly correspond to the most closer region to the binary where secular motion is allowed (at the right position of MMR 4/1). 

We check that dynamical maps not show difference in the closer regions if we choose initial conditions with $\Delta\varpi=0^o$ or with $\Delta\varpi=180^o$. We must note that in regions further from the binary, differences in structures appear inside the resonance as can see in Fig. 3 in \citet{Zoppetti.etal.2018}.

\begin{table}
\centering
\caption{Masses and orbital parameters of CB planets.}
\label{tab:CB}
\begin{threeparttable}

\begin{tabular}{|c|c|c|c|c|c|}
\hline
   System     &  $q$ & $\alpha$   & $e_B$   &  $m$ [$m_{\rm Jup}$] &   e    \\  
\hline
   Kepler-38$^{\;1}$  & 0.26 &   3.16     & 0.10 & 0.38                 &  0.03  \\  
   Kepler-453$^{\;2}$ & 0.21 &   4.26     & 0.05 & 0.05                 &  0.04  \\  
   TOI-1338$^{\;3}$   & 0.28 &   3.49     & 0.16 & 0.10                 &  0.09  \\  
\hline
\end{tabular}
$^{1}$ \citet{Orosz+2012K38}, $^{2}$ \citet{Welsh+2015}, $^{3}$ \citet{Kostov+2020}
\end{threeparttable}

\end{table}

\begin{figure*}[ht!]
\centering
\includegraphics[width=0.9\columnwidth,clip]{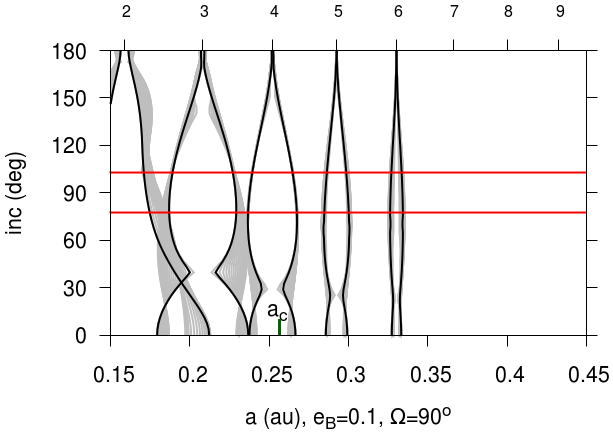} \includegraphics[width=0.9\columnwidth,clip]{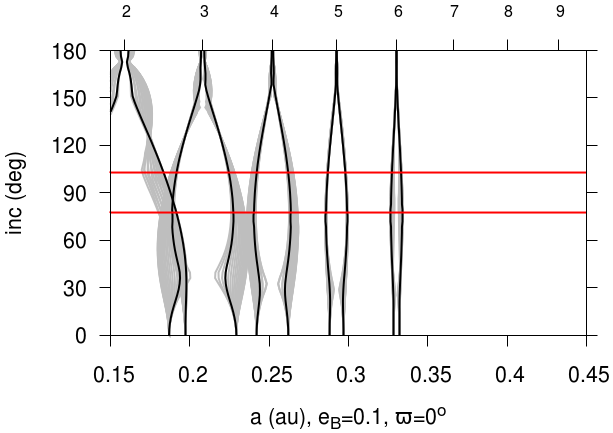} \\
\includegraphics[width=0.9\columnwidth,clip]{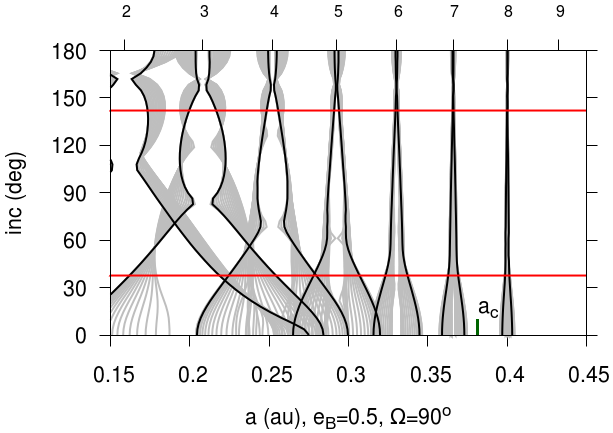} \includegraphics[width=0.9\columnwidth,clip]{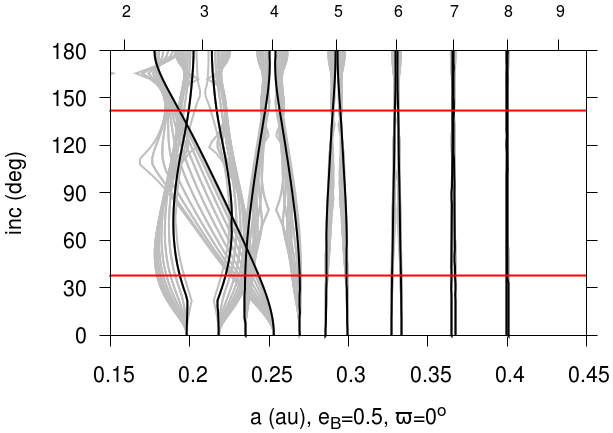} \\
\includegraphics[width=0.9\columnwidth,clip]{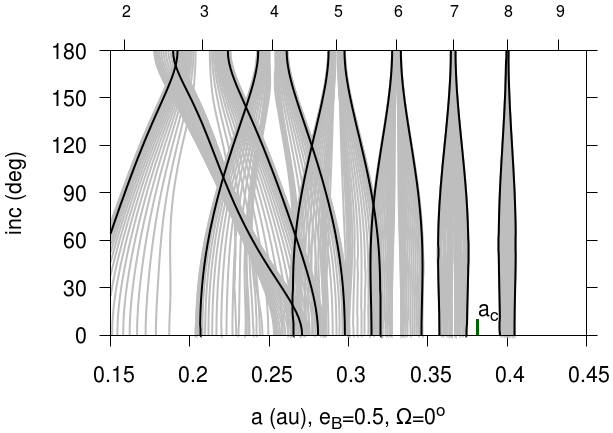} \includegraphics[width=0.9\columnwidth,clip]{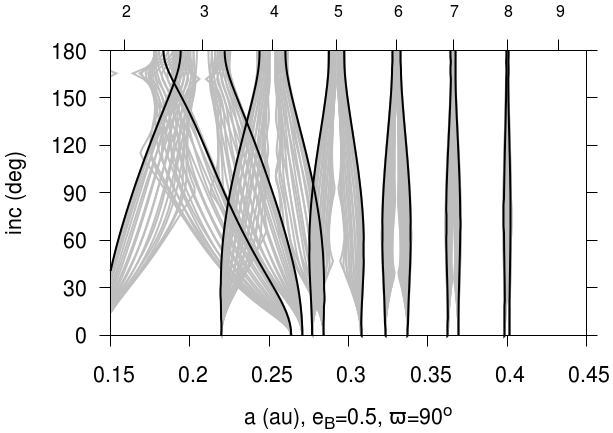} \\

\caption{Widths and location of N/1 resonances setting different values of $\varpi$ (left frames) and $\Omega$ (right frames) for a planet around a binary with mass ratio $q = 0.2$ and $a_B=0.1$ in the plane ($a,i$). Top scale of each frame indicates the positions of $N/1$ resonances with respect of the mean motion of the binary. We calculate each resonance with the semi-analytical model for 20 different values of $\varpi$ in left-hand frames ($\Omega$ in right-hand frames) from $0\degr$ to $180\degr$ and plotted each one with gray lines. Black lines denote the value for $\varpi=180\degr$ (left frames) or $\Omega=180\degr$ (right frames). Horizontal red lines indicates the separatrix location for nodal libration, thus regions where the dynamics only allow libration of $\Omega$ around $90\degr$. See text for more detail. Bottom panels do not present horizontal lines because the nodal resonance is not present, then $\Omega$ circulates independently of inclination $i$.}
\label{fig.wpn}
\end{figure*}

\subsection{Inclined circumbinary planets}\label{sec:cbp-inclined}
In the view of variety of CBDs around binaries and to test the model for arbitrary inclination, we studied the dynamics of inclined planets around binaries (P-type planets). We must note that for inclined particles, the nodal libration (those regions where inclination and node librates) is determined by the separatrix $F$, simplifying the expression of \cite{Farago&Laskar2010},
\begin{equation}\label{eq:F}
F = \sqrt{\frac{1 - e_{\rm B}^2}{1 - 5 \, e_{\rm B}^2\,\cos^2 \Omega + 4 \, e_{\rm B}^2}} \,\,\,.
\end{equation}
Then, libration of node, $\Omega$, and inclination, $i$ of CB planet around $90^o$ is expected when
\begin{equation}\label{eq:limit}
\arcsin F < i < \pi - \arcsin F \,\,\,.
\end{equation}
It is important also to remark that outside the separatrix region, $\Omega$ will circulates (prograde or retrograde direction). We recall that librating orbits and stationary angles change when brown dwarf companions are considered \citep{Chen+2019}, i.e. masses not compatible with CB planets detected. 

We explore the MMR location and widths around a wide variety of circumbinary inclined orbits for different binary mass ratios $q=\{0.2,0.5,1\}$. However, we choose to center our attention in three particular configurations for a mass ratio $q=0.2$. We use Eq. (\ref{dela2ja}) of semi-analytical model to calculate the width of the resonances, although we can search numerically the exact extension of each separatrix to estimate their interior and exterior limit. We only found that $2/1$ MMR exhibits appreciable differences between Eq. (\ref{dela2ja}) and the exact extension deduced from the level curves of the Hamiltonian. Although our planet is initially placed in circular orbits, due to the excitation induced by the binary, the mean value of the planet eccentricity in regular orbits is roughly $e\sim0.05$ and we use this value to calculate the width of the resonances.

The dynamics of CB inclined planets at these distances leads to circulation of $\varpi$ because the interaction with the binary. Also, outside the nodal resonances, the node $\Omega$ circulates \citep{Farago&Laskar2010,Doolin+2011, Li+2014}. Our model calculates the width of the resonance for a fixed value of $\varpi$ and $\Omega$ and we analyze the dependency of widths for different values from $0\degr$ to $180\degr$. We show the results in Fig. \ref{fig.wpn}, setting a binary with $q=0.2$. At left frames (right frames) we calculate the resonance width varying $\varpi$ ($\Omega$). Black lines indicate resonances setting $\varpi=180\degr$ (left frames) or $\Omega=180\degr$ (right frames). A common feature observed in all the panels is that:
\begin{itemize}
    \item the width of resonances decreases with increasing N.
    \item prograde resonances are wider than retrograde resonances.
    \item the changes in the width of the separatrix due to different values of $\varpi$ (left panels) is higher than when we change $\Omega$ (right panels). 
    \item wobble shape at high inclinations and for retrograde orbits is present when $\Omega=90^o$.
\end{itemize}
In each left-hand panel we indicate with $a_c$ the minimum stable coplanar-prograde semimajor axis, using the criteria of \citet{Holman+Wiegert1999}. This empirical criteria, widely used, is based in a polynomial fit that considers quadratical exponents of $a_B$, $e_B$ and $q$. According to our Figure \ref{fig.wpn} this limit roughly correspond to the most closer region to the binary where secular motion is allowed (outside the resonance overlap region).   

The difference between top and middle frames of Fig. \ref{fig.wpn} is the eccentricity of the binary.
For low eccentricity of the binary  ($e_B=0.1$ at top frames) setting different $\varpi$ values we observe the region of superposition in coplanar prograde orbits between 2/1 and 3/1 MMR, while the other resonances almost remain constant. The change of $\Omega$ at top right-frame does not affect coplanar prograde nor retrograde resonances.
When we increase the eccentricity of the binary ($e_B=0.5$, center frames), the separatrix at prograde coplanar resonances experiments huge changes, overlapping between them up to 6/1 MMR. Retrograde resonances are less affected and location of 4/1 MMR is associated to the most stable region near the binary. Island of stable motion are possible, where the gray lines does not cross at polar inclinations $i\sim90\degr$.

\begin{figure}[ht!]
\includegraphics[width=0.85\columnwidth,clip]{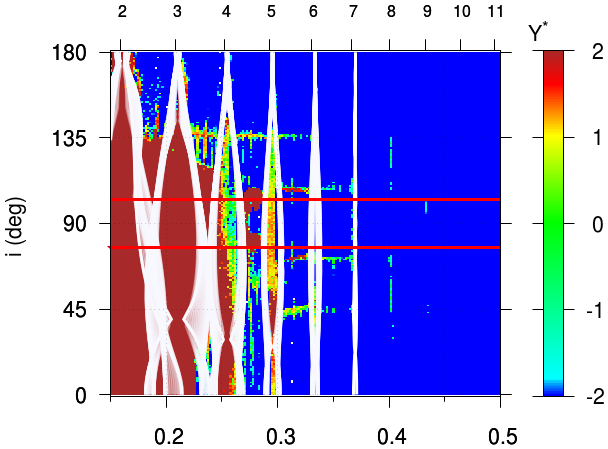} \\
\includegraphics[width=0.85\columnwidth,clip]{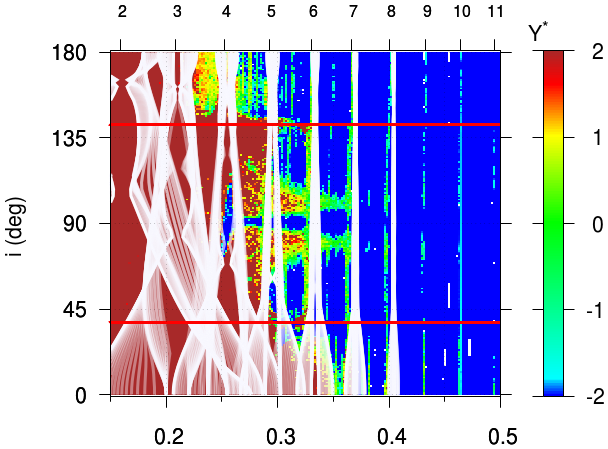} \\
\includegraphics[width=0.85\columnwidth,clip]{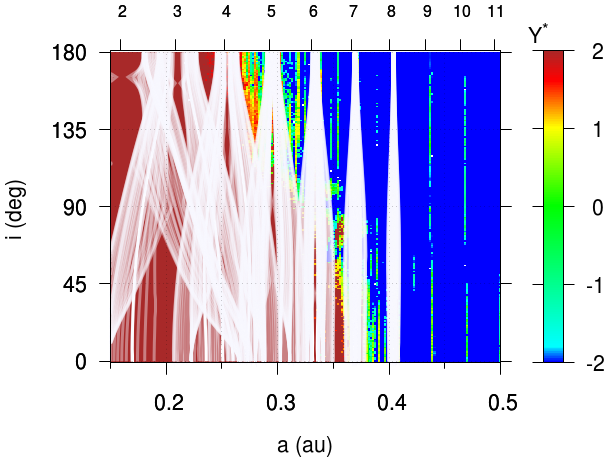} \\
  \vspace{-2em}
  \caption{Dynamical maps of inclined Jupiter planets around a binary with $(e_B, \Omega)=(0.1, 90\degr)$, $(e_B, \Omega)=(0.5, 90\degr)$, and $(e_B, \Omega)=(0.5, 0\degr)$ from top to bottom panels respectively. The planet has initial $\varpi=0\degr$, $e=0$, and the binary mass ratio is $q=0.2$. Horizontal red lines delimit the nodal separatrix using Eq. (\ref{eq:limit}). Top scale of each frame indicates the positions of $N/1$ resonances with respect of the mean motion of the binary. Each frame is constructed from the numerical integration of a grid of $300 \times 300$ initial conditions in the $(a,i)$ plane. All other initial orbital elements were set to the nominal values. Integration time was set to $3 \times 10^2$ years ($\sim 10^4$ orbital periods of the binary). White lines are the separatrices of each $N/1$ resonances using the semi-analytical model, varying the values of $\Omega$ and $\varpi$ (equivalent to gray lines in Fig \ref{fig.wpn}).}
\label{fig.binN90}
\end{figure}

Varying $\Omega$ (center right-frame) does not affect the width of low-inclined prograde resonances, but enhances width of highly-inclined retrograde resonances. At bottom left frame of Fig. \ref{fig.wpn} we plot the resonance widths for a fixed value of $\Omega=0\degr$ and varying $\varpi$, where the initial conditions are outside the libration region of $\Omega$. The separatrix location encompasses for wider regions in coplanar and highly inclined systems, while for retrograde coplanar orbits the separatrices shrinks and left favorable conditions for regular orbits closer to the binary. Bottom right frame shows separatrix extension for $\varpi=90\degr$ and different fixed values of $\Omega$, where inclined orbits are more affected by this variation.

Some works showed dynamical maps around binaries in the plane $(a,i)$ using ejection times \citep[see e.g][]{Doolin+2011} or Megno indicator \citep[see e.g][] {Cuello&Giuppone2019, Giuppone+Cuello.2019} but do not explain convincingly the spatial regular motion. 
To compare with the  model, we run numerical integrations and results are shown in Fig. \ref{fig.binN90}, using color scale $Y^*=\text{log}_{10}(\langle Y \rangle-2)$, where $\langle Y \rangle$ is the Megno indicator. We also overlay with white lines the $N/1$ resonances that we construct for the Fig. \ref{fig.wpn}.

Top frame of Fig. \ref{fig.binN90} shows $e_B=0.1$ (almost like Kepler-38, Kepler-453, or TOI-1338), where unstable regions shown in brown coincides with the overlap of $N/1$ resonances. While the prograde MMRs have considerable widths, as we increase the inclination the widths of MMR shrinks. The reported islands of stability around polar orbits \citep[see e.g][]{Doolin+2011, Cuello&Giuppone2019, Giuppone+Cuello.2019} are easily explained with the resonance width of MMR at highly inclined orbits ($i$ around $90^o$). Also is important to remark that as inclination increases, the width of resonances diminished, being retrograde resonances more compact.

When we increase the binary eccentricity $e_B=0.5$ (center and bottom frames of Fig. \ref{fig.binN90}, similar to Kepler-34), more chaotic regions near the binary at coplanar prograde and retrograde orbits appears as result of resonance overlap (see the center frames of Fig.  \ref{fig.wpn}). Regular orbits closer to the binary survives at $i\sim90\degr$. Finally at bottom frame of Fig. \ref{fig.binN90} we show the case were $e_B=0.5$ but dynamical maps with initial $\Omega=0\degr$ (means that $\Omega$ not fall inside the {nodal} libration region for any inclination and always circulates). Unstable regions of low-inclination prograde orbits appears that agree with the N/1 plotted resonances. It is necessary to recall that the circulation of $\varpi$ provokes overlap observed at bottom frame of Fig. \ref{fig.wpn}.

We must note that in the dynamical maps is evident a shift in semimajor axis for the location of nominal MMR. This shift is higher as we increase the mass ratio, $q$, or the binary eccentricity, $e_B$ (only of order of $1\%$ in the case of a binary with q=0.2 and $e_B=0.1$). This shift is also present using mean values of semimajor axis for dynamical maps.

\begin{figure}[ht!]
\centering
\includegraphics[width=\columnwidth,clip]{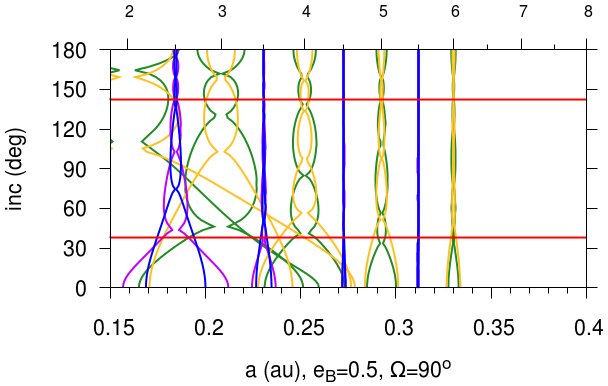} \\
\caption{Resonance $N/1$ and $N/2$ widths setting different mass ratio between components $q=\{0.2,1\}$ in the plane ($a,i$) calculated with the semi-analytical model. For $q=0.2$ we use green ($N/1$) and magenta ($N/2$) while for $q=1$ we use orange ($N/1$) and blue ($N/2$) resonances. Horizontal lines indicates the separatrix location for nodal libration. See the text for more detail.}
\label{fig.qrat}
\end{figure}

Finally, in Figure \ref{fig.qrat} we compare the width of MMR $N/1$ and $N/2$ around a binary with eccentricity $e_B=0.5$ in the plane ($a,i$) for different mass ratios $q=0.2$ and $q=1$. We set fixed values of $\Omega=90\degr$ and $\varpi=0\degr$. Mass ratio of binary components modify considerably the width of MMRs in prograde orbits (see e.g. $2/1$, $5/2$, $3/1$, and $4/1$). In prograde orbits, the resonance $2/1$ is the most important to determine instability regions because overlaps with regions of the resonance $4/1$ (or 5/1 depending on mass ratio).

Retrograde highly-inclined resonances almost not change their width with the mass ratio. For initial values of $\Omega=90\degr$ the wobble structure of resonances changes with $q$ and $i$; although is easy to identify regular motion islands beyond the position of $3/1$ resonance. Resonances $N/2$ are thinner than $N/1$; however due their location between two consecutive N/1 resonances, they overlap in regions that causes the chaos in retrograde configurations (see top axis in Figure \ref{fig.wpn}). Thus, the model is useful to interpret inclined unstable regions around binaries and we checked their validity under a wide combination of orbital parameters.

\section{Conclusions}\label{sec.conclusions}

This work presents a semi-analytical model for planetary mean-motion resonances 
around single or binary stars, yielding the location and stability of fixed 
points of the averaged system, libration period of small-amplitude oscillations, 
as well as widths of the libration region in the semimajor axis domain. It is a 
natural extension of the model developed by \cite{2020CeMDA.132....9G} for the 
restricted three-body problem.

Applications to the HD31527 and HD74156 planetary systems, as well as to fictitious and real
circumbinary planets, show a very good agreement between the model's predictions 
and dynamical features deduced from pure N-body simulations, even for very
high eccentricities. These dynamical characteristics include the widths of the
resonances, libration periods and stability of the fixed points. Together, these
features allow for a more detailed interpretation of the dynamical maps and help
estimate regions corresponding to long-term orbital stability.

A new orbital fit of the radial velocity data of HD31527 presented in this work shows
that the middle and outer planets are probably locked in a 16/3 mean-motion resonances
exhibiting a high-eccentricity moderate-amplitude libration of the resonant angle $16
\lambda_3 - 3\lambda_2 - 13\varpi_2$. This configuration appears to be stable for
timescales at least of the order of $10^7$ years; since no significant secular changes
in the semimajor axes or eccentricities are observed within this time frame, it is
possible that the orbital fit is actually stable for timescales comparable to the age
of the star. 

Differences between predicted and observed widths, as well as the location of the
resonance centers can be attributed to long-term variations in the secular angles
(i.e. longitudes of pericenter and ascending nodes) which are assumed constant in the
model. These effects are almost negligible in MMR among planetary bodies, but become
more visible in binary systems, especially for high mass-ratios between the stellar
components. Notwithstanding this effect, the general features of the resonant dynamics
are well reproduced by the model.

Using the semi-analytical model we convincingly demonstrated that for P-type orbits (circumbinary planets), the N/1 MMR overlap is directly associated with the instability limit in the regions closer to the binary. We confront our results with dynamical maps in both planes $(\alpha,e)$ and $(a,i)$.

When we analyze the MMR experimented by an inclined planet around a binary, we find that prograde resonances are wider than retrogrades, and N/1 resonance overlap is the main cause of instabilities near the binary. The resonances N/2 are considerably less stronger than N/1, but former contribute to overlap between consecutive N/1 resonances. We found that $N/1$ and $N/2$ resonances delimit the islands of stability for polar configurations and that resonance shape in the plane $(a,i)$ strongly depends on initial value of $\Omega$, either $0\degr$ or $90\degr$ for polar inclinations. As we move closer to the binary, the proximity causes circulation of $\Omega$ and $\varpi$ is short time-scales, thus the separatrix of each resonance is moving and creating a wider effective separatrix.

The model assumes that the center of the libration domain coincides with
the nominal position of the resonance. Although this approximation has been proved
very reliable in most of the cases, we have noticed some values of the system
parameters leading to a noticeable offset, appearing as a shift of the resonant
structure in the semimajor axis domain. While this issue is rarely observable in
planetary systems around single stars, it becomes more evident for circumbinary
planets with high values of the mass ratio $q$ or binary eccentricity, $e_B$. A more
detailed study of the resonance offset, as well as its description, is left for a
forthcoming work.

Codes for the computation of the resonances can be found at \href{http://www.fisica.edu.uy/~gallardo/atlas/plares.html}{www.fisica.edu.uy/$\sim$gallardo/atlas/plares.html}.

\begin{acknowledgements}
We thank the anonymous reviewer who, through a rigorous examination of the manuscript, suggested substantial improvements to this final version.
$N$-body computations were performed at Mulatona Cluster from CCAD-UNC, which is part of SNCAD-MinCyT, Argentina. T.G. acknowledges support from PEDECIBA. C.G. acknowledges the training acquired in the 2020 Sagan Exoplanet Summer Virtual Workshop.

\end{acknowledgements}

\bibliographystyle{aa}

\bibliography{respla}

\end{document}